\documentclass[11pt,a4paper]{article}
\usepackage{jcappub}
\usepackage{amsmath}
\usepackage{threeparttable}
\usepackage{booktabs}
 \newcommand{\D}{\mathrm{d}}

 \title{ Constraining    smoothness parameter and the  DD relation of Dyer-Roeder equation
         with supernovae }
\author[a]{Xi Yang,}
\author[a]{Hao-Ran Yu}
\author[a]{and Tong-Jie Zhang}
\affiliation[a]{Department of Astronomy, Beijing Normal University,\\ Beijing 100875, China}
\emailAdd{yangwds@mail.bnu.edu.cn} \emailAdd{yu@bnu.edu.cn} \emailAdd{tjzhang@bnu.edu.cn}

\abstract{ Our real universe is locally inhomogeneous. Dyer and Roeder introduced the smoothness parameter $\alpha$ to describe the influence of local inhomogeneity
on angular diameter  distance,  and they obtained the angular diameter  distance-redshift approximate relation
  (Dyer-Roeder equation) for locally inhomogeneous universe. Furthermore, the Distance-Duality (DD)  relation,
   $D_L(z)(1+z)^{-2}/D_A(z)=1$, should  be valid for all cosmological models that are described by   Riemannian
geometry,  where $D_L$ and $D_A$ are, respectively, the luminosity and
  angular distance distances. Therefore,  it is necessary  to test whether if the  Dyer-Roeder approximate equation  can satisfy the
  Distance-Duality  relation. In this paper, we use  Union2.1 SNe Ia data to constrain the smoothness parameter $\alpha$ and test whether  the Dyer-Roeder equation satisfies  the DD relation.  By using  $\chi^2$ minimization, we get
  $\alpha=0.92_{-0.32}^{+0.08}$ at $1\sigma$ and  $0.92_{-0.65}^{+0.08}$ at $2\sigma$, and our results show that the  Dyer-Roeder equation is in good consistency with  the DD relation at $1\sigma$.}

\keywords{supernova type Ia - standard candles, cosmological perturbation theory, weak gravitational lensing}

\begin{document}
\maketitle
 \flushbottom

%==================================introduction======================================

\section{Introduction}
In astronomy, the distance-redshift relation  is an important equation. The simplest approximation, cosmological principle, assumes that the universe
is homogeneous and isotropic. Using cosmological principle and Einstein equation,  the distance-redshift relation for homogeneous universe can be get \cite{wei2008}.    However, the real universe at least is locally inhomogeneous,
and the inhomogeneity can affect the distance-redshift relation. Therefore, many authors \cite{zel1964,ber1966,gun1967,kan1969} discussed this issue,
 but they did not give a simple formula for this relation. Until 1973,
 Dyer and Roeder \cite{dye1973} obtained a simple relation (named Dyer-Roeder (D-R) equation)
 via  the method of averaging path.

The D-R equation is an  angular diameter distance-redshift  approximate equation that contains a
parameter $\alpha$ (named smoothness parameter)
for locally inhomogeneous universe, where $\alpha$ ($0\leq \alpha \leq1$) describes the influence of local
inhomogeneity on the angular  diameter  distance.  The detailed  information of the D-R equation and
$\alpha$ are  displayed in Section \ref {DR}.

In order to  improve the  accuracy of the D-R equation, the value range of $\alpha$ need to be constrain.
Santos et al. \cite{sanet2008} used  the $182$ SNe Ia data of Riess et al. \cite{rie2007}  to obtain that $\alpha \geq 0.42$
 ($2\sigma$). Yu et al. \cite{ yu2011} got $\alpha=0.81_{-0.20}^{+0.19}$  at $1\sigma$ CL
via the observational $H(z)$ data, and Busti et al. \cite{bus2012} got $\alpha \geq 0.25$ ($2\sigma$) using Union2 SNe Ia data.
In this paper we use latest Union2.1 SNe Ia data which contains 580 SNe Ia  to constrain $\alpha$, and we get   $\alpha=0.92_{-0.32}^{+0.08}$  ($1\sigma$) and  $0.92_{-0.65}^{+0.08}$  ($2\sigma$).

 On the other hand, the distance-duality (DD) relation \cite{ eth1933,ell2007}
plays a fundamental  role in  modern cosmology, which reads
\begin{equation}  \label{1}
\frac{D_L}{D_A}(1+z)^{-2}  =  1,
\end{equation}
where $D_L$ and $D_A$ are, respectively, the luminosity and angular diameter  distances, and $z$ is the
 cosmological redshift. Doppler redshift and  redshift  of tired light theory \cite{wei2008}
 do not satisfy above expression.
This equation is valid for all cosmological models  based on  Riemannian
geometry if  photon travels through null geodesic and the photon number is conserved  \cite{ell1971},
so it should also be valid for an inhomogeneous universe  described by Riemannian geometry.
 The D-R equation is based on Robertson-Walker  metric and Einstein equations which are based on Riemannian
geometry (see Section \ref{DR}). So, on the surface, the D-R equation should satisfy the DD relation.   However,  the  D-R equation is  obtained via perturbation method,  and it just describes the relation of  angular diameter distance and redshift, which have nothing to
do directly with luminosity distance, therefore, it may put a  perturbation on the DD relation. So,
 whether if this equation  conflicts with  the DD relation need to  test. The D-R equation will be modified  if the DD relation cannot be satisfied.
   Fortunately,
 our results should that the D-R equation is in good consistency with the DD relation.
 In following  paragraph,  methods of testing the DD relation are  introduced.

%--------------------------------------------------------------------------------------------

A series of works have introduced the parameter $\eta(z)$ to tested the DD relation \cite{uza2004,hol2010,hol2011}, i.e.,
\begin{equation}   \label{2}
\frac{D_L}{D_A}(1+z)^{-2}=\eta(z),
\end{equation}
and there are generally two  ways to test it's validity.
The first way to constrain $\eta$ is to combine  the observed results of  $D_L$ and $ D_A$
both from observations \cite{bas2004,hol2010, li2011, men2012}.
This method is popular because it is cosmological model-independent. Holanda et al. \cite{hol2010}
artificially assumed  that $ \eta(z)$ takes two forms, i.e., $\eta(z)=1+\eta_0z$ or
$ \eta(z)=1+\eta_0z/(1+z)$, and got $\eta_0=-0.28^{+0.44}_{-0.44}$ ($2\sigma$ CL)
for the $D_A$ samples of  elliptical model \cite{def2005}. Therefore, their results just satisfy
the DD relation at $ 2\sigma$ CL for elliptical model. In subsequent papers, Li et al. \cite{li2011} and Meng et al. \cite{men2012}
 obtained the conclusion that DD relation can be accommodated  at the $1\sigma$ CL  for the elliptical model.
 In short, these  cosmological model-independent tests suggest that the deviation of $\eta$ from $1$ is minor.

%-----------------------------------------------------------------------------------------------------------

Therefore, these conclusions can at least  denote that  the   observed redshift  is mainly generated
by the expansion of universe, because other kinds of redshift do not satisfy Equation (\ref{1}).
So these results can strictly rule out some exotic cosmological models. For example, tired
light theory \cite{wei2008} assumes  that our universe is basically static
 and  the photons just suffer a loss of energy while they are traveling to us,
   so the redshift naturally increasing with the distance.
   This theory  concludes that  $D_L(1+z)^{-1/2}/D_A=1$.  Lima et al. \cite{lim2011} proposed
   that the small deviation of $\eta$ may indicate that some breaks on fundamental physical theories.

%----------------------------------------------------------------------------------------------------------

The second way to test DD relation is to combine the observational results of $D_L$ (or $D_A$) with
 theoretical results of $D_A$ (or $D_L$) for a given cosmological model \cite{uza2004,deb2006,avg2010,hol2011}.
Obviously, analyses obtained from this way are just valid for the given cosmological model, so this method have too much limitation.

%-------------------------------------------------------------------------------------------------------

    In this paper, we adopt the second method to constrain  $\eta$ and $\alpha$,
     i.e., we combine the observed values of $D_L(z)$  from supernovae data with
     the theoretical values of  $ D_A(z)$
from  Dyer-Roeder (D-R) equation to constrain $\alpha$ and $\eta$. Because in this way, our aim can be achieved.
Our aim is to constrain    $\alpha$ and test whether if the D-R equation satisfies the DD relation.

%----------------------------------------------------------------------------------------------------------------

%-----------------------------------

This paper is organized as follows: In Section \ref{DR}, we present the D-R equation.
 Section \ref{sam} briefly introduces the sample of SNe Ia data: Union2.1.
  The analysis methods and results are presented in Section \ref{s3}.
   Finally  we  give the discussions and conclusions  in Section \ref{s5}.
%==========================introduction over====================================================================
%==========================introduction over======================================

%===========================section two===========================================
\section{Dyer-Roeder equation} \label{DR}
%------------------------------------------------------------------------------------------------------------------

   We consider a light ray which comes to us from a far object�� propagating along a tube of density ($\rho_{\rm int}=\alpha\rho_0$) in a background homogeneous universe of density $\rho_0$.
     According to optical scalar equations \cite{sac1961}, the angular diameter
     distance $D_A$ satisfies the following formula:
\begin{equation}   \label{3}
 \frac{\D^{2}D_A}{\D s^{2}}=-\left(
 |\sigma|^{2}+\frac{1}{2}R_{\rm \alpha \beta}k^{\alpha}k^{\beta}
 \right),
\end{equation}
 where $s$, $\sigma$, $k^{\alpha}$, and $R_{\rm \alpha \beta}$ are, respectively, the affine parameter,
 the shear of the light bundle, the vector tangent to the light ray, and the Ricci tensor. For symmetry, the shear $\sigma=0$.
Combining the Robertson-Walker  metric  and Einstein  equations, one can
 obtain the distance-relation equation (D-R equation) for a locally inhomogeneous and flat-space $\Lambda$CDM cosmological model:
\begin{subequations}
\begin{equation} \label{xx}
\frac{\D^{2}D_A}{\D z^{2}} + \mathcal{P}\frac{\D D_A}{\D z} + \mathcal{Q}{D_A} = 0,
\end{equation}
where,
\begin{eqnarray}
  \mathcal{P} =\frac{\frac{7}{2} \Omega_M (1+z)^3 +2(1-\Omega_M)}{\Omega_M (1+z)^4 +(1-\Omega_M)(1+z)},\\
 \mathcal{Q} = \frac{ \frac{3}{2}  \alpha \Omega_M  }{\Omega_M (1+z)^2 + (1-\Omega_M)(1+z)^{-1}}.
 \end{eqnarray}
 Here the initial conditions are
 \begin{eqnarray}
\begin{cases}        \label{ini}
  D_A(0)=0,\\
 \frac{ \D D_A}{\D z} \big|_{z=0} =\frac{c}{H_0},\\
 \end{cases}
 \end{eqnarray}
 \end{subequations}
 where $c$ and $H_0$ are the speed of light and Hubble constant respectively.
 The smooth parameter $\alpha$  is defined as
\begin{equation}
\alpha=\frac{\rho_{\rm int}}{\rho_0},
\end{equation}
where $ \rho_{\rm int}$ is the mean density of  intergalactic matter in the universe, while
$\rho_{0}$ is the mean density of the whole universe, so $ \alpha \in [0,1]$. When $\alpha=0$,
it means  all the matter clustered into stars, galaxies, while $\alpha=1$ means the universe
is totally homogeneous. Therefore, the smooth parameter should satisfy $0<\alpha < 1$ in the real universe,
and $D_A$ is a
decreasing function of $\alpha$ \cite{lin1988}.
 Obviously, the smooth parameter should be different at various epoch of
 the universe \citep{sanet2008}. The functional forms of $\alpha(z)$ have been discussed
  in Ref. \citep{lin1988,lin1998, ras2009,san2008,bol2011}. However,
   we will regard $\alpha$ as a constant   because there are no convinced functional forms for $\alpha(z)$.
   This equation  have no analytical solution \cite{kan1998,kan2000,kan2001},  but one can get its approximate
solution \citep{dem2003}.

 %--------------------------------------------------------------------------------------------------------------------

%=====================================================================
\section{Samples} \label{sam}
We choose 580 SNe Ia data  of Union2.1\footnote[1]{\url{http://supernova.lbl.gov/Union}} \cite{suz2012}
  which cover the redshift interval $0.015\leq z\leq 1.414$,
   to constrain the parameters $\alpha$ and $\eta$.
  The Union2.1  compilation is an updated version of  Union2 \cite{ama2010} compilation by adding new SNe data
  from the Hubble Space Telescope Cluster Survey to Union2  compilation.
  All 580 SNe of Union2.1  were fit using the  light curve fitter SALT2
  \cite{guy07}.

      The SALT2 fitter  fits all SNe Ia with three parameters. The three parameters are $m_{B}^{\rm max}$,
       $x_1$,
       and $c_1$. $m_{B}^{\rm max}$ is the integrated B-band flux at maximum light.
        $x_1$ is the deviation from the average light-curve
    shape,  and $c_1$ is the deviation from the mean SNe Ia B--V color.
     The linear combination of the above parameters forms the empirical formula of  distance modulus:
\begin{equation}
\mu_B=m_{B}^{\rm max}+\alpha x_1-\beta c_1 -M_B,
\end{equation}
where $\alpha$ and $\beta$  are constant coefficients,  and $M_B$ is absolute B-band
magnitude of an SNe Ia.   $m_{B}^{\rm max}$,  $x_1$,    and $c_1$ are got by fitting the light curves of SNe Ia.
Therefore, the distance modulus  fitted by SALT2  have three unknown parameters $\alpha$, $\beta$, and $M_B$,
 i.e., $\mu=\mu(\alpha,  \beta, M_B)$.
  The best fitted values of  $\alpha$, $\beta$,  and $M_B$ are got by $\chi^2$-minimization method with a
  given cosmological model, and then, the best fitted values of  distance modulus $\mu(\alpha,  \beta, M_B)$ is got.
  It is important  to note that the distance modulus $\mu$ of SNe Ia from Union2.1 is dependent on Hubble constant $H_0$,
  because  the   given cosmological model has $H_0$, which will bring $H_0$ into $M_B$ if one use $\chi^2$-minimization.
  In Union2.1, they chose $H_0= 70  \rm km s^{-1} Mpc ^{-1}$ to get $M_{B}$ and $\mu$
  (see Equation (4) and Table 6 in Ref. \cite{suz2012}). So, in this paper,
  we must choose $H_0= 70  \rm km s^{-1} Mpc ^{-1}$  in D-R equation to constrain  $\alpha$ and $\eta$.
%===========================section three================================================
\section{Analysis  and results}  \label{s3}

%-----------------------------------------------------------------------------------------------------------------

 Firstly, we  get  the numerical solution of $D_A$ from Equation (\ref{xx}):
 \begin{subequations}
 \begin{equation}
 D_A = D_A(H_o, \Omega_M , \alpha; z).
 \end{equation}
 Combining  Equation (\ref{2}) with above expression, we  obtain
\begin{equation}   \label{b}
 D_L=\eta(1+z)^2D_A(H_o, \Omega_M, \alpha;z) = D_L(H_o, \Omega_M, \alpha , \eta; z).
 \end{equation}
 The distance modulus-luminosity distance relation is
 \begin{equation}         \label{gg}
 \mu = -5 +5\lg{D_L},
 \end{equation}
 where $u$ is distance modulus.
 Substituting Equation (\ref{b}) into Equation (\ref{gg}),  we  get
 \begin{equation}
 \mu=\mu(H_o, \Omega_M, \alpha, \eta;z).   \label{u}
 \end{equation}
 \end{subequations}
%-------------------------------------------------------------------------------------------------------------------

 In order to constrain $\alpha$ and $\eta$, we use $ \chi^2$ minimization and plot the contours  on  the $(\alpha, \eta)$
 plane. The contours of the confidence level of $68.3\%$ and $95.4\%$ are
 determined by two parameter levels $2.30$ and $6.18$ respectively.
  For each parameter, the confidence of
  $68.3\% (1\sigma)$ and $95.4\% (2\sigma)$,  are
  determined by one parameter levels $1.00$ and $4.00$.
 The formula of $\chi^2$ minimization is
 \begin{equation}
 \chi^2(H_0, \Omega_M, \alpha, \eta) = \sum_i
 \bigg [
 \frac{\mu(H_0, \Omega_M, \alpha, \eta;z_i)-\mu_{\rm obs}(z_i)}{\sigma (z_i)}
  \bigg ]^2  \label{chi2}
  \end{equation}
  where $u(H_0, \Omega_M, \alpha, \eta; z)$ is obtained from Equation (\ref{u}), and $u_{\rm obs}$ is
  the observational value of SNe Ia with error $\sigma (z_i)$  in Union2.1.

%-----------------------------------------------------------------------------------------------------------------
 In the analysis, we marginalize  $\Omega_M$  by integrating over them,
 and select $H_0= 70  \rm km s^{-1} Mpc ^{-1}$.
 There are some things need to be emphasized. Firstly, there is no Gaussian prior on $\Omega_M$, and we just assume that $\Omega_M$
 has a uniform distribution in interval $[0, 1]$.  Secondly, we  must select  $H_0= 70  \rm km s^{-1} Mpc ^{-1}$ in our analysis,
     because the distance modulus of  Union2.1 compilation is dependent on $H_0$, and the  Union2.1 sample assumed
     $H_0= 70  \rm km s^{-1} Mpc ^{-1}$ to get the distance modulus of SNe Ia (see Section \ref{sam}).

     We assume that $\eta$ is a constant, i.e.,
     \begin{equation}
     \eta=1+\delta, \label{delta}
     \end{equation}
      where $\delta$ is a little constant. When $\delta=0$, the DD relation is satisfied.
\begin{figure}[htb]
\begin{center}
\includegraphics[width=0.56 \textwidth]{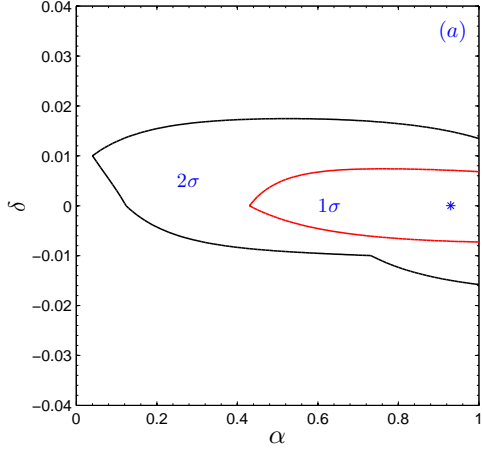}
  \end{center}
  \caption{$(a)$: Confidence regions at $68.3\%$  and $95.4\%$ levels from
  inner to outer respectively on the $(\alpha, \delta)$ plane  for the sample Union2.1. The $``*"$ in the center
  of confidence regions indicates the best fitted values $(0.93,0)$.}
  \label{fig1}
\end{figure}
\begin{figure}[htb]
\begin{center}
$\begin{array}{ccc}
\includegraphics[width=0.53\textwidth,height=0.33\textwidth]{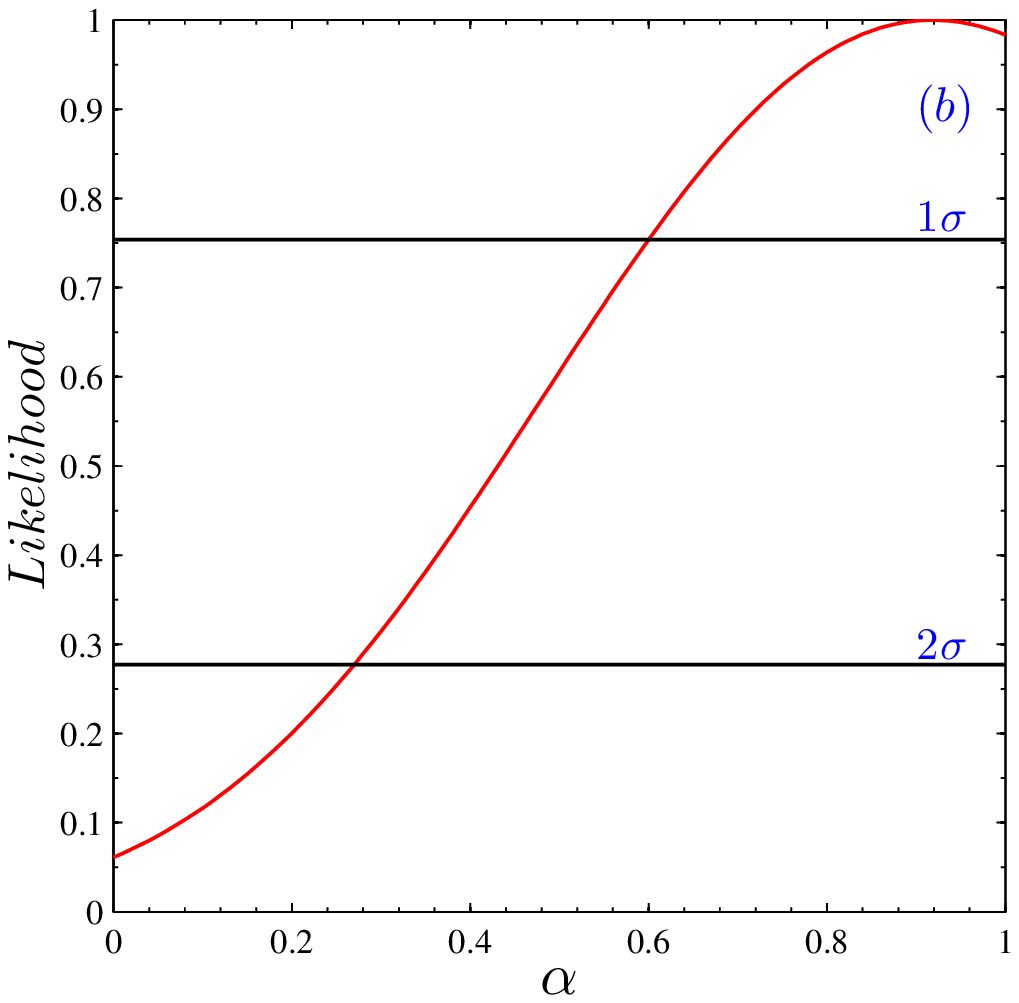}&
\includegraphics[width=0.53\textwidth,height=0.33\textwidth]{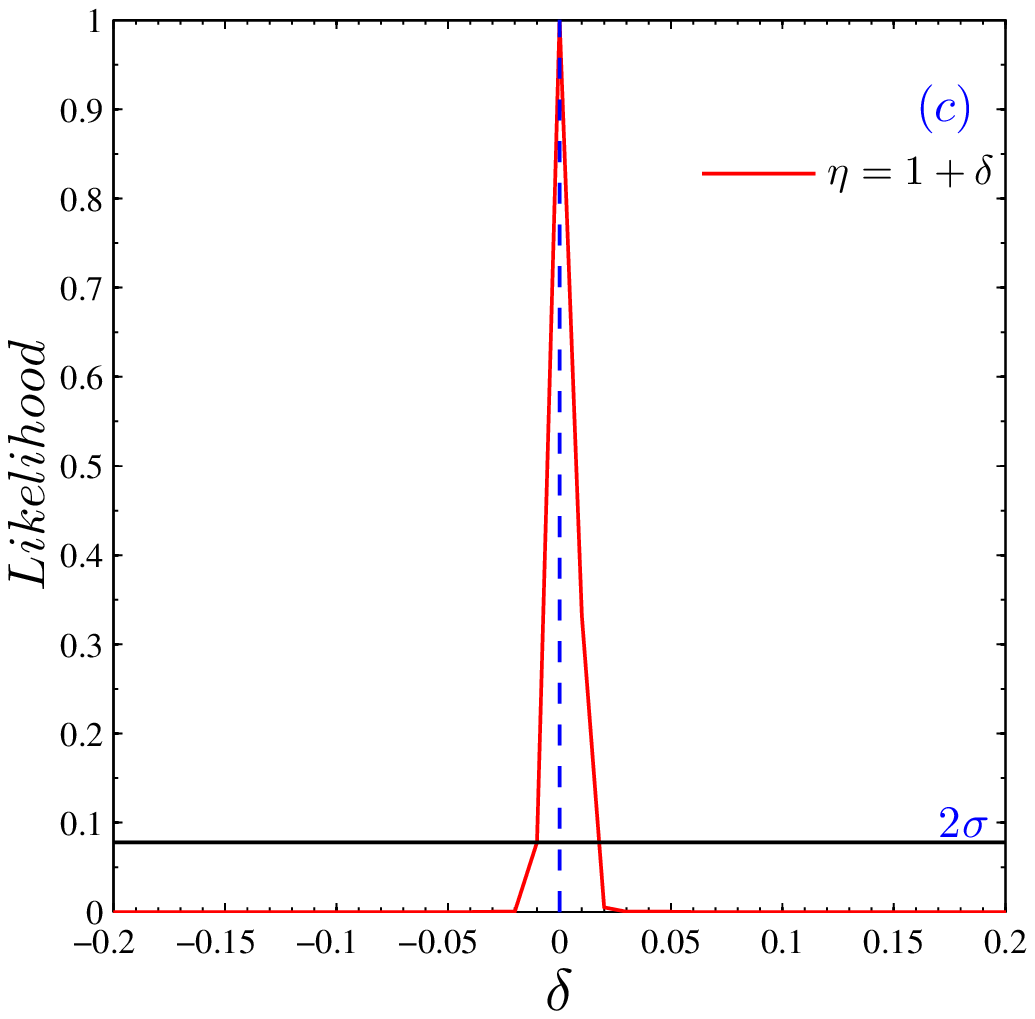}
\end{array}$
\end{center}
\caption{$(b)$: PDF for  $\alpha$. $(c)$: PDF for  $\delta$. Obviously, the D-R equation is in good consistency with the DD relation ($\delta=0$). The step length of $\delta$ is $0.01$ in our analysis, which is larger than the  range of values for  $\delta$ at $1 \sigma$, so we do not plot $1 \sigma $ confidence interval of $\delta$ on the picture.}
\label{fig2}
\end{figure}

     By marginalizing $\Omega_M$,  we obtain that  the best values of $\alpha$ and $\delta$ are $0.93$ and $0$ respectively. The two-dimensional confidence regions of $\alpha$ and $\delta$ are shown in Figure \ref{fig1}(a).
     In Figure \ref{fig2} (b) and (c), the one-dimensional probability distribution of $\alpha$ and $\delta$ are plotted
     respectively.  Figure \ref{fig2}(b) shows that
      $\alpha=0.92_{-0.32}^{+0.08}$ at $1\sigma$ and  $0.92_{-0.65}^{+0.08}$ at $2\sigma$.  For Figure \ref{fig2}(c), the result is $\delta=0_{-0.01}^{+0.01}$ at $2 \sigma$.  The step length of $\delta$ is $0.01$ in our analysis, which is larger than the  range of values for  $\delta$ at $1 \sigma$, so we do not give  $1 \sigma $ confidence interval of  $\delta$.  Because  the DD relation is valid  when $\delta=0$, these results display that
      the D-R approximation equation is in consistency with the DD relation.  For a inhomogeneous universe which is
      described by Riemannian geometry, the DD relation should be met. Therefore,
        the D-R approximation equation is well if we  use the DD relation to weigh it.

%%%%%%%%%%%%%%%%%%%%%%%%%%%%%%%%%%%%%%%%%%%%%%%
%----------------------------------------------------------------------------------------------------------------
%======================section four==========================================================
\section{Discussions and Conclusions} \label{s5}
With the improvement of precision
 of observations in astronomy, accurate estimate of the distance is requested.  Therefore,
  many  theories which describe the inhomogeneity  have been proposed to correct the distance-redshift relation \cite{bol20112}.
Such as, LTB model \cite{cel2000}, swiss-cheese model \cite{kan1969},  weak-lensing   approximation \cite{pyn2004}, and Dyer-Roeder relation \cite{dye1972}.  Among them,  the Dyer-Roeder equation is the simplest distance estimator, because it is obtained by adding small perturbation on FLRW model.   Therefore,   constraining $\alpha$ from D-R equation is essential.

In this article, we use the latest Union2.1 SNe Ia sample to constrain $\alpha$. We find that
$\alpha\geq0.60$ at $1 \sigma$, $\alpha\geq0.27$ at $2 \sigma$, and the best fitted value of $\alpha$ is $0.92$.
The D-R equation will reduce to general distance-redshift relation of $\Lambda$CDM model if $\alpha=1$,
so our results show that the distance-redshift equation of $\Lambda$CDM model is not the best equation for distance.
Furthermore,  the D-R approximate equation  is a perturbation equation.  Whether it  put a disturbance on the DD relation
 need to test.   Our results illustrate that the D-R equation does not violate the DD relation.
So the D-R approximation equation  is well if we  use the DD relation to weigh it.

 Of course,  $\alpha$ should not be a constant, but  a function of $z$ \citep{sanet2008}. However, Current data is not
  enough  to study its functional forms $\alpha(z)$.
  Bolejko. K \cite{bol2011} studied the relation between
 D-R  approximation equation and weak-lensing   approximation,
 and they concluded that the two approximations  could be
  compatible when $[\alpha(z)-1] \sim 1/(1+z)^{5/4}$. In the future, with the increase of  high-redshift SNe Ia data,
the   approximate functional forms of $\alpha(z)$ may be determined, and then,
the estimate of  distance will be more accuracy.

%=================section four over===================================%
\acknowledgments
X. Y. is very grateful to Jian-Chuan Zheng, Ji-Long Chen, Zhong-Xu Zhai, Cong Ma  and  Shuo Yuan  for their kind help. This work was supported by the National Science
Foundation of China (Grants No. 11173006), the Ministry
of Science and Technology National Basic Science program
(project 973) under grant No. 2012CB821804, and the Fundamental
Research Funds for the Central Universities.

%=========================================================

%==================== bibliography===============================================
%==================== bibliography===============================================


\begin{thebibliography}{99}
\bibitem{wei2008}
  S. Weinberg,   \emph{Cosmology,}
   Oxford University Press, Oxford  (2008)

   \bibitem{ber1966}
 B. Bertotti,  \emph{The Luminosity of Distant Galaxies,}
 Proc. R. Soc. London A  \textbf{294} (1966) 195


\bibitem{gun1967}
  J. E. Gunn,  \emph{On the Propagation of Light in Inhomogeneous Cosmologies. I. Mean Effects,}
   Astrophys. J.   \textbf{150} (1967)  737


\bibitem{kan1969}
 R.  Kantowski, \emph{Corrections in the Luminosity-Redshift Relations of the Homogeneous Fried-Mann Models,}
 Astrophys. J.     \textbf{155} (1969)  89

 \bibitem{zel1964}
 Ya. B. Zeldovich, \emph{Observations in a Universe Homogeneous in the Mean,}
 Soviet Astronomy \textbf{8} (1964) 13

 \bibitem{dye1973}
  C. C.  Dyer and   R. C. Roeder, \emph{Distance-Redshift Relations for Universes with Some Intergalactic Medium,}
   Astrophys. J.   \textbf{180} (1973)  L31

 \bibitem{sanet2008}
 R. C.  Santos, J. V. Cunha and  J. A. S. Lima,  \emph{Constraining the dark energy and smoothness-parameter with supernovae,}
 Phys. Rev. D  \textbf{77} (2008) 023519    [arXiv:0709.3679v2]


\bibitem{rie2007}
 A. G. Riess,  L. G. Strolger, S. Casertano  et al.,
 \emph{New Hubble Space Telescope Discoveries of Type Ia Supernovae at$z > 1$: Narrowing Constraints on the Early Behavior of Dark Energy,}
 Astrophys. J.     \textbf{659} (2007)  98 [arXiv:astro-ph/0611572v2]

 \bibitem{yu2011}
H. R. Yu,  T. Lan,  H. L.  Wan,  T. J.  Zhang  and  B. Q.  Wang, \emph{Constraints on smoothness parameter and dark energy using observational $H(z)$ data,}
 Research in Astron. Astrophys.    \textbf{11} (2011)  125  [arXiv:1008.1935v1]

\bibitem{bus2012}
V. c. Busti,  R. C. Santos and J. A. S. Lima,  \emph{Constraining the dark energy and smooth parameter with type Ia
supernovae and Gamma-Ray Bursts,} Phys. Rev. D  \textbf{85} (2012) 103503 [arXiv:1202.0449]

\bibitem{ell2007}
  G. F. R.  Ellis, \emph{On the definition of distance in general relativity: I. M. H. Etherington,}
Gen. Rel. Grav. \textbf{39} (2007) 1047

\bibitem{eth1933}
 I. M. H. Etherington,  \emph{On the Definition of Distance in General Relativity,}
 Phi. Mag. \textbf{15}  {1933} 761

\bibitem{ell1971}
G. F. R. Ellis, \emph{Republication of: Relativistic cosmology,}
Gen. Rel. Grav.  \textbf{41}  (2009) 581

\bibitem{hol2010}
 R. F. L. Holanda, J. A. Lima  and  M. B. Ribeiro,  \emph{Testing the Distance-Duality Relation with Galaxy Clusters and Type Ia Supernovae,}
    Astrophys. J.     \textbf{722} (2010)  L233 [arXiv:1005.4458v2]

\bibitem{hol2011}
 R. F. L. Holanda, J. A. Lima  and  M. B. Ribeiro,  \emph{Cosmic Distance Duality Relation and the Shape of Galaxy Clusters,}
 Astron. Astrophys.   \textbf{528} (2011)  L14 [arXiv:1003.5906v2]

 \bibitem{uza2004}
 J. P. Uzan,  N.  Aghanim  and  Y.  Mellier, \emph{The distance duality relation from X-ray and SZ observations of clusters,}
 Phys. Rev. D    \textbf{70} (2004) 083533    [arXiv:astro-ph/0405620v1]

\bibitem{bas2004}
 B. A. Basset and  M. Kunz, \emph{Cosmic distance-duality as probe of exotic physics and acceleration,}
 Phys.Rev. D \textbf{69} (2004) 101305 [arXiv:astro-ph/0312443v2]

 \bibitem{li2011}
 Z. Li,  P. Wu and H. Yu,  \emph{Cosmological-model-independent tests for the distance-duality relation from Galaxy Clusters and Type Ia Supernova,}
  Astrophys. J.     \textbf{729} (2011)  L14  [arXiv:1101.5255v2]

 \bibitem{men2012}
X. L. Meng,  T. J. Zhang,  H. Zhan  and  X. Wang, \emph{Morphology of Galaxy Clusters: A Cosmological Model-Independent Test of the Cosmic Distance-Duality Relation,}
Astrophys. J.     \textbf{745} (2012)  98 [arXiv:1104.2833v2]

\bibitem{def2005}
E. De Fillips,  M. Sereno, M. W. Bautz  and  G. Longo, \emph{Measuring the Three-Dimensional Structure of Galaxy Clusters. I. Application to a Sample of 25 Clusters,}
Astrophys. J. \textbf{625} (2005) 108 [arXiv:astro-ph/0502153v1]


\bibitem{lim2011}
 J. A. S. Lima, J. V.  Cunha  and   V. T. Zanchin, \emph{Deformed Distance Duality Relations and Supernovae Dimming,}
  Astrophys. J.     \textbf{742} (2011)  L26 [arXiv:1110.5065v1]

\bibitem{avg2010}
A. Avgoustidis, C. Burrage, J. Redondo, L. Verde. and  R. Jimenez,  \emph{Constraints on cosmic opacity and beyond the standard model physics from cosmological distance measurements,}
JCAP \textbf{10} (2010) 024 [arXiv:1004.2053v1]

\bibitem{deb2006}
F. De Berbardis,  E. Giusarma  and  A. Melchiorri, \emph{Constraints on Distance Duality Relation from Sunyaev Zel'dovich Effect and Chandra X-ray measurements,}
Int. J. Mod. Phys. D  \textbf{15} (2006) 759 [arXiv:gr-qc/0606029v1]

   \bibitem{sac1961}
  R.  Sachs, \emph{Gravitational Waves in General Relativity. VI. The Outgoing Radiation Condition,}
  Royal Society of London Proceedings Series A   \textbf{264} (1961) 309

   \bibitem{lin1988}
   E. V.  Linder,  \emph{Light propagation in generalized Friedmann universes,}
   Astron. Astrophys.   \textbf{206} (1988)  190

\bibitem{bol2011}
 K. Bolejko, \emph{Weak lensing and the Dyer-Roeder approximation,}
 MNRAS \textbf{412} (2011) 1937  [arXiv:1011.3876v1]


\bibitem{lin1998}
E. V.  Linder,  \emph{Transition from Clumpy to Smooth Angular Diameter Distances,}
Astrophys. J.     \textbf{497} (1998)  28 [arXiv:astro-ph/9707349v2]

\bibitem{ras2009}
 S.  R\"{a}s\"{a}nen, \emph{Light propagation in statistically homogeneous and isotropic dust universes,}
JCAP     \textbf{02} (2009)  011 [arXiv:0812.2872v2]

\bibitem{san2008}
 R. C.  Santos  and J. A. S.  Lima,  \emph{Clustering, Angular Size and Dark Energy,}
 Phys. Rev. D  \textbf{77} (2008) 083505    [arXiv:0803.1865v1]

 \bibitem{kan1998}
R. Kantowski,  \emph{The Effects of Inhomogeneities on Evaluating the mass parameter $\Omega_m$ and the cosmological constant $\Lambda$,}
Astrophys. J.     \textbf{507} (1998)  483  [arXiv:astro-ph/9802208v2]

\bibitem{kan2000}
  R. Kantowski, J. K.  Kao  and R. C. Thomas,  \emph{Distance-Redshift Relations in Inhomogeneous Friedmann-Lemaitre-Robertson-Walker Cosmology,}
  Astrophys. J.     \textbf{545} (2000)  549

\bibitem{kan2001}
 R. Kantowski and  R. C. Thomas, \emph{Distance-Redshift in Inhomogeneous $\Omega_0=1$ Friedmann-Lemaitre-Robertson-Walker Cosmology,}
  Astrophys. J.     \textbf{561} (2001)  491  [arXiv:astro-ph/0011176v2]

\bibitem{dem2003}
 M. Demianski, R. de Ritis, A. A. Marino and  E. Piedipalumbo, \emph{Approximate angular diameter distance in a locally inhomogeneous universe with nonzero cosmological constant,}
Astron. Astrophys. \textbf{411} (2003) 33  [arXiv:astro-ph/0310830v1]

 \bibitem{suz2012}
   N.   Suzuki,  D. Rubin,   C. Lidman  et al., \emph{The Hubble Space Telescope Cluster Supernova Survey: V. Improving the Dark Energy Constraints Above $z>1$ and Building an Early-Type-Hosted Supernova Sample,}
   Astrophys. J.    \textbf{746} (2012)  85  [arXiv:1105.3470v1]

\bibitem{ama2010}
  R. Amanullah, C. Lidman, D. Rubin et al.,\emph{Spectra and Hubble Space Telescope Light Curves of Six Type Ia Supernovae at $0.511 < z < 1.12$ and the Union2 Compilation,}
  Astrophys. J. \textbf{716} (2010) 712 [arXiv:1004.1711v1]

\bibitem{guy07}
  J. Guy, P.  Astier,  D. Hardin  et al.,  \emph{SALT2: using distant supernovae to improve the use of Type Ia supernovae as distance indicators,}
  Astron. Astrophys.   \textbf{466} (2007)  11 [arXiv:astro-ph/0701828v1]

\bibitem{bol20112}
K. Bolejko,  M. N.  C$\acute{e}$l$\acute{e}$rier and A.  Krasi$\acute{n}$ski,
\emph{Inhomogeneous cosmological models: exact solutions
and their applications,}  Class. Quantum Grav.  \textbf{28} (2011)  164002  [arXiv:1102.1449]

\bibitem{cel2000}
M. N.  C$\acute{e}$l$\acute{e}$rier, \emph{Do we really see a cosmological constant in the supernovae data?}
Astron. Astrophys.   \textbf{353} (2000)  63  [arXiv:astro-ph/9907206]

\bibitem{pyn2004}
T.  Pyne  and   M.  Birkinshaw,  \emph{
The luminosity distance in perturbed FLRW space-times,}
MNRAS  \textbf{348} (2004)  581  [arXiv:astro-ph/0310841]



\bibitem{dye1972}
 C. C. Dyer and    R. C. Roeder, \emph{The Distance-Redshift Relation for Universes with no Intergalactic Medium,}
  Astrophys. J.   \textbf{174} (1972)  L115

\end{thebibliography}
\end{document}